\documentclass[reprint,amsmath,amssymb,aps,prl,groupedaddress,nofootinbib,superscriptaddress]{revtex4-1}
\usepackage{graphicx,xcolor,tikz}
\usepackage{amsthm,amssymb,amsmath,dsfont,braket}
\usepackage{bm}
\usepackage[hidelinks,pagebackref=false,pdfnewwindow=true]{hyperref} 
\usepackage{epstopdf,psfrag}
\usepackage{relsize,amsbsy}
\usepackage[export]{adjustbox}

\newcommand{\1}{\mathds{1}}

\newcommand{\ms}{Majorana spinon }
\newcommand{\mss}{Majorana spinons }

\newcommand{\Hs}[1]{\mbox{$\mathcal{H}$-$#1$}}
\newcommand{\zt}{\mathbb{Z}_2}

\begin{document}
	
	\title{Raman scattering in correlated thin films as a probe of chargeless surface states
	}
	\author{Brent Perreault}
	\affiliation{\small School of Physics and Astronomy, University of Minnesota, Minneapolis, Minnesota 55455, USA}
	%
	\author{Johannes Knolle}
	\affiliation{\small Department of Physics, Cavendish Laboratory, JJ Thomson Avenue, Cambridge CB3 0HE, U.K.}
	\author{Natalia B. Perkins}
	\author{F. J. Burnell}
	\affiliation{\small School of Physics and Astronomy, University of Minnesota, Minneapolis, Minnesota 55455, USA}
	
	\date{\today} 
	
	\begin{abstract}	
		Several powerful techniques exist to detect topologically protected surface states of weakly-interacting electronic systems. In contrast, surface modes of strongly interacting systems which do not carry electric charge are much harder to detect. We propose resonant light scattering as a means of probing the chargeless surface modes of interacting quantum spin systems, and illustrate its efficacy by a concrete calculation for the 3D hyperhoneycomb Kitaev quantum spin liquid phase. We show that resonant scattering is required to efficiently couple to this model's sublattice polarized surface modes, comprised of emergent Majorana fermions that result from spin fractionalization. We demonstrate that the low-energy response is dominated by the surface contribution for thin films, allowing identification and characterization of emergent topological band structures.
	\end{abstract}
	
	\maketitle		
	
	
	{\it Introduction.} 
	One of the most striking recent developments in condensed matter physics has been the discovery that certain types of three-dimensional (3D) band structures harbor {\it topologically protected surface states}, which cannot be gapped out without breaking a bulk symmetry. Systems with such surface states include topological insulators~\cite{FuKaneMele,MooreBalents,Roy}, Weyl semimetals~\cite{balents11,wan11,vafek14,yang14,potter14,lv15}, and a number of others~\cite{matsuura13,chen15}.   
	In these weakly-interacting systems where the quasiparticles carry electric charge, theoretical predictions have quickly led to experimental detection: the Dirac cone surface states of 3D topological insulators were first detected several years ago~\cite{Hsieh08,Chen09,Xia09} using high-resolution angle-resolved photoemission spectroscopy (ARPES). More recently, ARPES has also detected the Fermi arcs characteristic of Weyl semimetals in compounds TaAs, TaP, NbAs and NbP~\cite{xu15-1,xu15-2,huang15,lee15-4}.
	
	Such surface states are not restricted to weakly-interacting electronic systems. In fact, topological surface states have been predicted in a number of Mott-insulating systems where 
	they often originate from spin fractionalization in quantum spin liquids (QSLs) \cite{pesin10,ZhangRan10,schaffer15,hermanns15-1}. 
	This intriguing possibility poses an experimental challenge: since surface probes such as ARPES and STM couple to charge, then how can such chargeless surface states be detected?
	For bulk properties of chargeless topological states, much progress has been made recently in probing candidate QSLs using inelastic neutron~\cite{Coldea01,Coldea,Helton07,Helton10,Han,Fak2012,punk14,knolle14-1,knolle15,smith15} and Raman~\cite{Lemmens03,ko10,Wulferding,wulferding2,knolle14-2,Gupta,Sandilands,perreault15} scattering. Both measurements couple to spin degrees of freedom (d.o.f.), and hence can give signatures of fractionalized {\it spinon} excitations.  
	
	\begin{figure}
		\centering
		\includegraphics[width=.82\linewidth,valign=t]{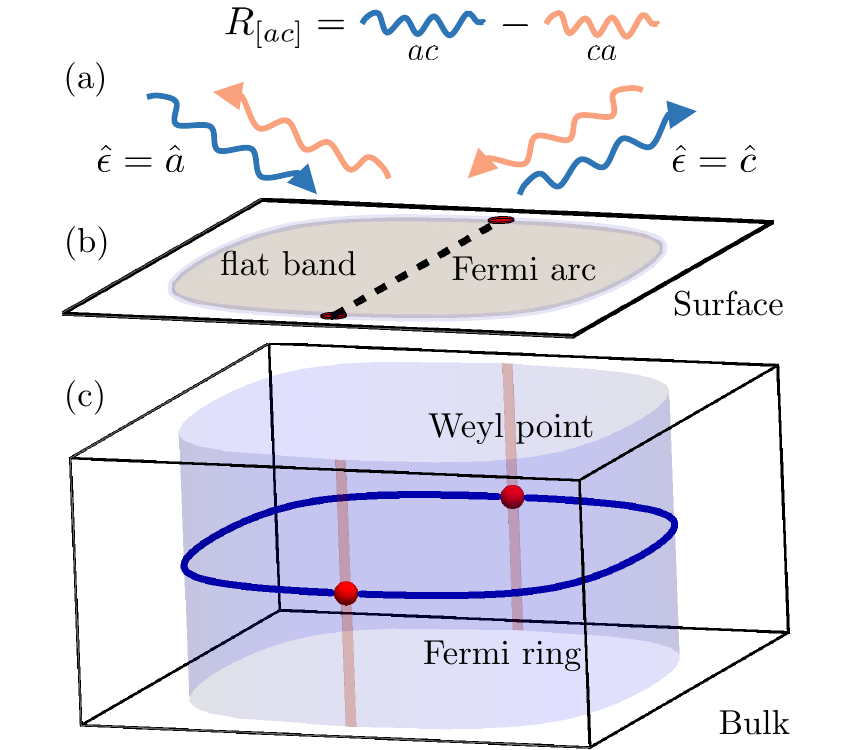}
		\caption{(c) Schematic of the Fermi ring and limiting positions of the Weyl points of the emergent chargeless Majorana fermions as $\kappa \to 0$ in the bulk Brillouin Zone (BZ). (b) the projection of the Weyl points onto the surface BZ and the surface flat band ($\kappa =0$) and Fermi arc ($\kappa >0$). (a) illustrates a resonant light-scattering process in the anti-symmetric channel that can probe the surface modes.}
		\label{Fig1}
	\end{figure}  
	
	Here we study inelastic light scattering as a probe of the chargeless topological surface states that can arise in strongly-interacting systems.  We focus on the example of the Kitaev QSL on the hyperhoneycomb lattice \cite{mandal09,lee14}, which is known to harbor such boundary states \cite{schaffer15,matsuura13}. This model is of particular interest because the insulating magnet $\beta-$Li$_2$IrO$_3$ \cite{takayama15,modic14} is believed to be described by an effective Hamiltonian on a hyperhoneycomb lattice with dominant Kitaev-type interactions  \cite{jackeli09,sizyuk14,rau14,kim15,biffin14-1,biffin14-2,lee15,lee15-2,kimchi14}.  Our main findings are that (1) the surface modes can be identified by considering the low-energy power laws in spectra of thin films; and (2) the surface states contribute significantly to the light-scattering response only in the resonant regime~\cite{shastry90,shastry91,ko10}.  

	Our proposal is summarized schematically in Fig.~\ref{Fig1}. The idea is that though light scattering is typically a bulk probe, when applied to sufficiently thin films  
the surface responses can be observable if the density of states (DOS) of surface states decays more slowly with frequency than the bulk DOS so that the surface response dominates at sufficiently low energies.  We show that this occurs for both time-reversal (TR) symmetric and TR broken\cite{hermanns15-1} topological phases of the hyperhoneycomb Kitaev QSL, allowing direct experimental detection of the corresponding topological surface states. 


	{\it 3D Kitaev model.} 
	The Kitaev Hamiltonian~\cite{kitaev06} is 
	\begin{align}\label{H1}
	{H}_K &= J \sum_{\left<ij\right>_\alpha} \sigma^\alpha_i \sigma^\alpha_j \ \ ,
	\end{align}
	where $\left<ij\right>_\alpha$ are nearest-neighbor (NN) bonds, $\sigma^\alpha_j$ are the Pauli matrices, and
	$\alpha =\{x,y,z\}$ specifies which components of spins interact along each of three inequivalent bonds. The model is solved exactly by replacing the spin operator at each site $j$ with four Majorana fermions $c_j$ and $b_j^\alpha$ via $\sigma^\alpha_j =i b_j^\alpha  c_j$~\cite{kitaev06}. In terms of Majorana fermions, the Hamiltonian in Eq.~(\ref{H1}) takes the form 
	${H}_K=J\sum_{\langle ij\rangle_\alpha}u_{{\langle ij\rangle}^\alpha}c_ic_j$,  where the $u_{{\langle ij\rangle}^\alpha}\equiv ib_i^\alpha b_j^\alpha$
	form a $\zt$ lattice gauge field for the $c_j$ Majoranas. The $u_{{\langle ij\rangle}^\alpha}$ commute with each other and the Hamiltonian, and are therefore static. 
	
	In the Majorana description, the physical d.o.f are the fluxes of the $\zt$ gauge theory on elementary plaquettes of the lattice, and dispersing \mss $c_i$ in the flux background. The ground state on a given lattice corresponds to a fixed flux configuration,
	which is flux free for the \Hs{0} lattice~\cite{mandal09,obrien15,kimchi14}.  
	
	\begin{figure}
		\centering
		\includegraphics[width=0.8\linewidth,valign=t]{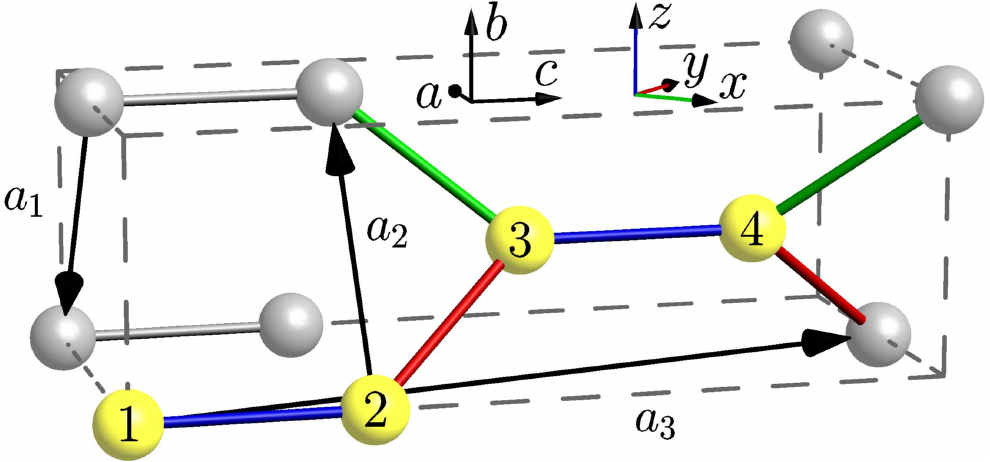}
		\caption{The hyperhoneycomb lattice, \Hs{0}.  The lattice vectors are
			$\mathnormal{ a}_{1/2} = (-1,\mp\sqrt{2},0)$ and
			$\mathnormal{ a}_3 = (-1,0,3) $.}
		\label{Fig2}
	\end{figure}  
	
	In this flux-free configuration,
	the Hamiltonian is quadratic in the \mss~$\{c_i \}$.  Diagonalization leads to a band structure with two distinct bands on the \Hs{0} lattice, where the modes at zero energy form a one-dimensional Fermi ring shown schematically by the blue line in Fig.~\ref{Fig1}(c).  With open boundaries, surface flat bands occur within the projection of the Fermi ring onto the surface BZ (see Fig.~\ref{Fig1}(b))~\cite{schaffer15}.  This band structure and the associated surface states are protected by a combination $\mathcal{C}= I \mathcal{T}$ of inversion and TR symmetry, which is a sublattice symmetry within the \ms description~\cite{schaffer15}. The gapless surface modes are sublattice polarized, and hence protected from back-scattering~\cite{matsuura13}.
	
	Applying a magnetic field $H_h=\sum_{j,\alpha}{h^\alpha}S^\alpha_j$, where $\{h^x,h^y,h^z\}$ are all nonzero, destroys the Fermi ring by breaking TR, and therefore the sublattice symmetry $\mathcal{C}$. 
	If $h$ is much smaller than the flux gap $\Delta$, the low-energy Hamiltonian retains the zero-flux ground-state flux configuration, and the first non-vanishing TR symmetry-breaking terms appear at third order in $h$ \cite{kitaev06}. 
	The relevant term at low-energy is
	\begin{align} \label{2Hops}
	H_h =  \kappa  \sum_{\langle\langle ij \rangle\rangle_\gamma} 
	\sigma^{\alpha}_i \sigma^\gamma_{l} \sigma^{\beta}_j
	= i \kappa \sum_{\langle\langle ij \rangle\rangle_\gamma} \tilde{u}_{\langle\langle ij \rangle\rangle_\gamma}  c_i c_j,	
	\end{align} 
	where $ (il), (jl)$ are pairs of neighbors along a bond of type $\alpha$ and $\beta$ respectively, and $\gamma$ is complementary. In terms of \mss this gives a next-nearest neighbor (NNN) hopping term 
	with $\tilde{u}_{\langle\langle ij \rangle\rangle_\gamma} \equiv u_{\langle il\rangle_\alpha} u_{\langle lj \rangle_\beta}$  
	and $\kappa \sim h^x h^y h^z / \Delta^2$. 
	On the \Hs{0} lattice, the magnetic field perturbation gaps out the Fermi ring, leaving a pair of  Weyl points~\cite{hermanns15-1}, which are fixed to the Fermi energy by the unbroken inversion symmetry~\cite{obrien15}.
	The surface flat bands are reduced to surface Fermi-arcs connecting the projection of the Weyl points in the surface Brillouin zone (BZ)~\cite{wan11} (see Fig.~\ref{Fig1}).    
	
	\begin{figure}
		\centering
		\includegraphics[width=.82\linewidth,valign=t]{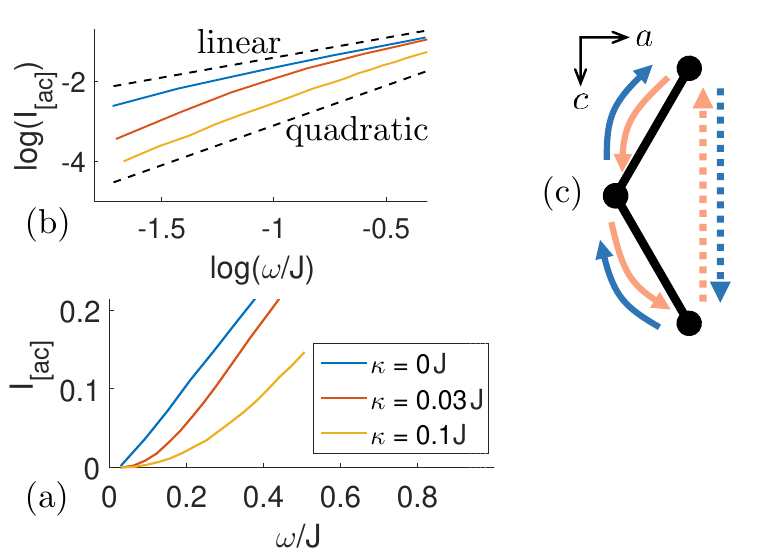}
		\caption{(a) Low-energy bulk scattering intensity in the anti-symmetric channel $I_{[ac]}$ 
			for various values of the effective perturbation $\kappa$ and (b) the log-log version. The index $[ac]$ represents the channel antisymmetrized over ``in" and ``out" polarizations in the $a$ and $c$ directions \cite{supp}. (c) illustrates resonant light-scattering processes contributing to the [ac] channel, which mimics the DOS'.  
		}
		\label{Fig3}
	\end{figure} 
	
	{\it Raman scattering. } 
	The key features of the bulk and surface \ms bands described above can be detected using inelastic photon spectroscopy. 
	To establish this, we first review some important aspects of the derivation of the Raman operator in Mott insulators~\cite{shastry90,shastry91,ko10} to show that the magnetic field has a negligible effect, and clarify the resonant processes of interest.
	
	Most inelastic light scatting of low-energy magnetism has been in the regime of Raman spectroscopy \cite{Devereaux07,hayes12}, which probes excitations ranging 1--100 meV (10--1000cm$^{-1}$) \cite{polian03,hayes12}. However, given the expected Kitaev exchange-interaction scale ($J$) in the honeycomb iridates of around 2 -- 4 meV \cite{Vamshi14,sizyuk14,katukuri16}, the energy scales discussed here are in the regime of Brillouin scattering: 0.01--1 meV (0.1--10 cm$^{-1}$) \cite{polian03,hayes12}, which differs from Raman only by the spectrometer. We continue to refer to `Raman' operators and spectra although each applies for both experiments.  
	
	At zero temperature the Raman response  
	can be written as a correlation function of scattering operators
	\begin{align}\label{I}
	I(\omega) = 2\pi \int d\omega e^{i\omega t}\braket{R(t)^\dagger R(0)},
	\end{align}
	where $\omega=\omega_{\rm in}-\omega_{\rm out}$ is the total energy transferred from the in- and out-going photons to the system, and in the following we assume that $\omega\ll \omega_{\rm in (out)}$.  
	For a Mott-insulator, the Raman operator is 
	\begin{align}\label{R1}
	R= -P H_t^{
		{\boldsymbol \epsilon}_{\rm out}^*
	} (H-i\eta)^{-1} H_t^{
	{\boldsymbol \epsilon}_{\rm in}} P,
\end{align}
where $P$ is the projector onto states with a fixed electron occupancy per site, ${\boldsymbol \epsilon}_{\rm in}$ and $ {\boldsymbol \epsilon}_{\rm out}$ are the incoming and outgoing photon polarization vectors, respectively, and $H_t^{\epsilon}$ is the electron/photon vertex for the polarization $\epsilon$ given by
\begin{align}
H_t^{\boldsymbol{\epsilon}} = \left(\frac{i e}{\hbar c}\right)\sum_{n,n',\gamma,\gamma'} (\mathbf{d}_{n,n'}\cdot \boldsymbol{\epsilon}) T^{\gamma,\gamma'}_{n,n'} a^\dagger_{n,\gamma} a_{n',\gamma'}.
\end{align}
Here $T^{\gamma,\gamma'}_{n,n'}$ describes the electronic hopping matrix,  $\mathbf{d}_{n,n'}$ is the spatial vector from lattice site $n$ to site $n'$, and $a_{n,\gamma}$ is the annihilation operator for an electron at site $n$ with $\gamma$ running over spin and orbitals. 

The full Hamiltonian in the resolvent $(-H+i\eta)^{-1}$ can be written as $H=H_t+H_U$, where $H_t$ is the electronic hopping Hamiltonian, and for convenience we define
the interaction term $H_U$ describing the on-site electron interactions, such as Coulomb repulsion and Hund's coupling, relative to the initial photon energy: $H_U=H_{\text{int}} - \omega_{\rm in}$. The resolvent $(-H+i\eta)^{-1}$ can be formally expanded to give
\begin{align}\label{RT}
R= P H_t^{\boldsymbol{\epsilon}_{\text{out}}^*} \left[H_U^{-1} + H_U^{-1}H_t H_U^{-1} + ... \right] H_t^{\boldsymbol{\epsilon}_{\text{in}}} P,
\end{align}
(we dropped $-i\eta$).
In the presence of a magnetic field, the resolvent in Eq.~(\ref{RT}) has an additional small parameter  proportional to $h/U$:
\begin{align}
\left[H_U + H_t + H_h\right]^{-1} = H_U^{-1}\left[\1 + H_t H_U^{-1} + H_h H_U^{-1} + ... \right].
\nonumber
\end{align}  
Hence, in the regime $h \ll t$ we can neglect the magnetic field during the scattering process. 

If $t/(U-\omega_{\rm in})$ is small, electron hopping is strongly suppressed, and the derivation of the Raman operator proceeds as it does for a spin-exchange Hamiltonian. The lowest-order terms contributing to $R$ are linear in $t/(U-\omega_{\rm in})$ and have the well-known Loudon-Fleury (LF) form~\cite{fleury68}
\begin{align}\label{RF}
R& = \sum_{n,n';\alpha,\beta} (\mathbf{d}_{n,n'}\cdot \boldsymbol{\epsilon}_{\text{in}}) (\mathbf{d}_{n,n'}\cdot \boldsymbol{\epsilon}_{\text{out}}^*) H_{n,n'}^{\alpha,\beta}\sigma^\alpha_n \sigma^\beta_{n'} ,
\end{align}
where $H_{n,n'}^{\alpha,\beta}$ defines the generic spin-exchange Hamiltonian on the bonds $\langle n,n'\rangle$, which we take as the pure Kitaev model. 

For NN Kitaev interactions, the LF vertex does not couple to fluxes, probing only \ms band structures~\cite{knolle14-2}. However, because the NN spin-exchange processes that enter into (\ref{RF}) involve both sublattices, \emph{the LF vertex cannot couple to the sublattice-polarized surface flat bands}. Moreover, even if  the sublattice symmetry is broken with a small magnetic field, the surface states are still approximately polarized, and the LF vertex couples only very weakly to the surface states.  

In order to detect the gapless surface modes, a Raman operator should contain terms coupling two \mss on the same sublattice. 
Such a term can appear by tuning the photon frequency resonant with the Mott gap so that $t/(U-\omega_{\rm in})$ is no longer very small, and higher-order terms in the expansion of Eq.~(\ref{RT}) contribute intermediate electron hops.  Two such resonant hopping processes, involving three different sites and an NNN electronic hop, are illustrated in Fig.~\ref{Fig3}(c); such processes can lead to the desired low-energy term:
\begin{align}\label{Rres}
R_{\text{res}} &= i \kappa' \sum_{\ll ij \gg^\gamma}  
\sigma^{\alpha}_i \sigma^\gamma_{l} \sigma^{\beta}_j  \times A_{ilj} \nonumber \\ &
= -\kappa' \sum_{\langle\langle ij \rangle\rangle_\gamma} \tilde{u}_{\langle\langle ij \rangle\rangle_\gamma}  c_i c_j  \times A_{ilj}
\end{align} 
where $\kappa'$ contains the electron-photon coupling and spin-exchange constants. There are other three-spin terms with $\alpha,\beta,$ and $\gamma$ permuted in Eq.~(\ref{Rres}) that create fluxes and are therefore unimportant at low energies. 
The computation of the resonant light-scattering matrix elements follows Refs. \onlinecite{shastry91} and \onlinecite{ko10}; details for the iridates will be presented elsewhere.

\begin{figure*}
	\centering
	\includegraphics[width=\linewidth,valign=t]{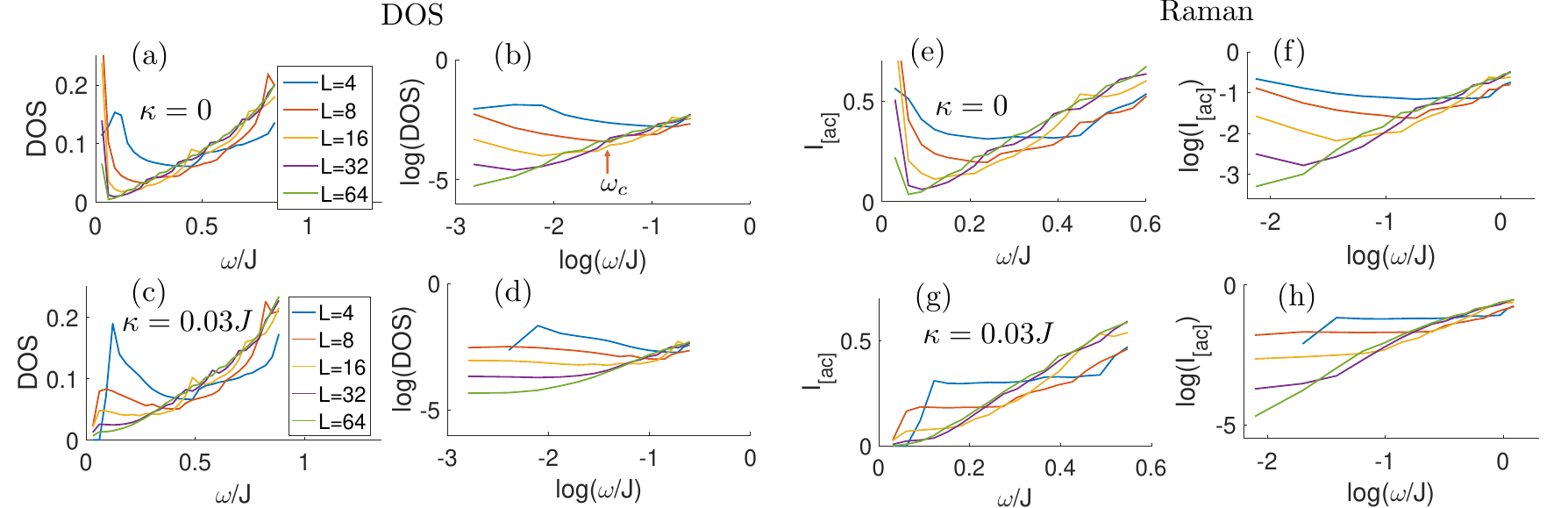}
	\caption{The low-energy DOS is plotted for (a) $\kappa=0$ and (c) $\kappa=0.03J$ for different slab widths $L$, measured in unit cells in the $a_1$ direction.  The low-energy peaks [plateaus] in the DOS are due to the surface flat bands in (a) [surface Fermi-arcs in (c)].
		(b) and (d) show the corresponding log-log plots illustrating the crossover between power laws describing the surface contribution to ones describing the bulk. (e), (f), (g), and (h) are the same plots for
		the resonant scattering intensity $I_{[ac]}$ in the antisymmetric $[ac]$ channel. The suppression of low-frequency modes in (g) compared to (c) is due to suppression by the scattering matrix elements. 
	}
	\label{Fig4}
\end{figure*}

Unlike the LF scattering vertex, the polarization-dependent factor $A_{ilj}=\left[\left(\boldsymbol{\epsilon}_{\text{in}}\cdot\mathbf{d}_{ji}\right) \left(\boldsymbol{\epsilon}_{\text{out}}^*\cdot\mathbf{d}_{il}\right)-\left(\boldsymbol{\epsilon}_{\text{out}}^*\cdot\mathbf{d}_{ji}\right) \left(\boldsymbol{\epsilon}_{\text{in}}\cdot\mathbf{d}_{il}\right)\right]$, is only non-zero in polarization channels that are anti-symmetric in the exchange of in and out polarizations.
This requires the use of polarization channels that do not appear with only the LF operator Eq.~(\ref{RF}). We focus on one of these, the anti-symmetric $[ac]$ combination of the two-photon processes (illustrated in Fig.~\ref{Fig1}), in which
one photon has polarization along $a$ and another along $c$. Due to the low symmetry of the model, isolating this channel requires observation of several directly-observable spectra. However, in the resonant regime the anti-symmetric part is expected to dominate the response in a few directly-observable channels, such as $I_{ab}$ \cite{supp}.
Because of the form of Eq.~(\ref{Rres}), the computation of low-energy Raman spectra for this system amounts to evaluating four-spinon dynamic correlators of a quadratic fermionic Hamiltonian \cite{knolle14-2,perreault15}. 

{\it Results.} 
The bulk light-scattering intensity at low frequency is shown for different values of $\kappa$ in Figs.~\ref{Fig3}(a) and \ref{Fig3}(b). The scattering intensity is almost linear at $\kappa=0$ and close to quadratic at $ \kappa=0.03J$ \cite{hfoot} directly reflecting power law changes in the \ms DOS anticipated in Ref.~\cite{hermanns15-1}. 

To study the Brillouin/Raman scattering response of the surface modes, we consider systems that are infinite in two directions but have a finite number $L$ of unit cells along the stacking direction $a_1$. This is one of several cutting planes with similar flat bands. However, cutting planes whose normal vector lies in the plane of $\mathbf{b}$ and $\mathnormal{ a}_{3}-\mathnormal{a}_{1}-\mathnormal{a}_{2}$ do not receive a projection from the interior of the bulk Fermi ring and therefore are not required to host a flat band.


Figs. 4(a) and (b) show the low-frequency behavior of the DOS in the unperturbed model ($\kappa=0$) in a linear and log scale, respectively. 
The low-energy peak in the DOS seen for different slab thicknesses indicates the presence of the flat band. The position of this peak is not strictly at zero frequency due to the top and bottom surface modes hybridizing, shifting it to higher energies. The peak drifts towards zero frequency for larger $L$, but its height relative to the rest of the spectrum decreases due to decreasing surface-to-bulk ratio. At low energies our results are consistent with the power law $-1$ expected for the surface states, in contrast to the linear power law seen above the crossover frequency $\omega_c$. Figs. 4(e) and (f) show the scattering intensity in the [ac] channel for $\kappa=0$, where we find that the behavior of the resonant Raman intensity reflects the behavior of the DOS, as expected. 

When $\kappa \ne 0$ most of the surface states are gapped, leaving only a surface Fermi arc at very low-energies, whose contribution to the DOS tends to a constant at zero frequency, in contrast with the quadratic power law seen above $\omega_c$. In practice, this constant behavior is dwarfed by the leftover peak from the flat bands, which have hybridized due to the symmetry-breaking perturbation. Hence the difference between surface flat bands and surface Fermi arcs is only detectable at energy scales below $\kappa$.
We plot the DOS for $\kappa=0.03J$ in Figs. 4(c) and (d). This value is below the local flux gap of $\Delta = 0.13J$ \cite{obrien15,kimchi14,smith15} putting it in the perturbative regime. We find that the flat band peaks are significantly suppressed upon introducing the magnetic field, although the low-energy power-laws in this case are not visible at the numerical resolution used of about $0.03J$. Similar behavior is observed in the Raman intensity shown in Figs. 4(g) and (h), up to some additional suppression of the flat-band peak due to matrix-element effects.

Experimentally, we expect that the low-energy peaks and power laws in the absence of a magnetic field should be discernible by Brillouin scattering for a film of thickness 20 or 30 unit cells or less, corresponding to 100 to 300nm in $\beta-$Li$_2$IrO$_3$.

{\it Discussion.} 
We have shown that low-frequency resonant light-scattering is a powerful probe of the QSL state, and can reveal both the bulk and surface \ms DOS in the 3D Kitaev QSL. For the hyperhoneycomb lattice, we find a bulk signature of the $\mathcal{T}$-broken Weyl spin liquid that arises upon perturbing the Kitaev Hamiltonian with a weak magnetic field, as well as a signature of topological surface states (both with and without $\mathcal{T}$-breaking) in thin films.  Our main results are that (1) the Weyl spin liquid can be distinguished from the parent $\mathcal{T}$-symmetric state by the power law governing the low-frequency response; (2) the symmetry properties of the usual LF vertex do not allow it to couple to sub-lattice polarized low-energy surface modes; and (3) the resonant Raman operator arising from the three-spin interaction can be used to probe the system's topologically-protected surface states on thin films.

Besides being able to couple to the surface flat bands, the anti-symmetric channels facilitate the task of separating the low-frequency response of the QSL from the contributions of acoustic phonons and Rayleigh scattering \cite{Rousseau,Klein}. 
Specifically, phonons contribute to antisymmetric channels only if they couple to resonant electron hops~\cite{Rousseau,Klein}; hence these processes will be suppressed. Rayleigh scattering leaves the state unchanged after a single two-photon event so there is no difference between two polarization combinations. 

Brillouin and Raman scattering on thin films are expected to be useful probes of surface modes in any other system with a large surface DOS relative to the bulk, which is typically the case for topological and symmetry-protected surface states. Hence, we expect that electronic Weyl semimetals can also be probed by light
scattering in addition to conventional ARPES and STM techniques. 
The vanishing coupling between the non-resonant scattering operator and the surface modes is specific to Mott insulators with sublattice-polarized surface modes. Nonetheless, the same symmetries appear in the Kitaev model on a few other lattices including the harmonic honeycomb series~\cite{modic14,schaffer15}, 
and the $(8,3)c$ lattice~\cite{obrien15}.  Further studies on lattices with different symmetry combinations~\cite{obrien15} are left for future work.

In addition, topological surface states without electric charge can also appear in non-fractionalized systems, e.g. from topological magnon bands in kagome ferromagnets \cite{katsura10}. Some of their bulk properties have been experimentally verified very recently \cite{hirschberger15,chisnell15} and we predict that the accompanying topological magnon surface states can be identified in a similar fashion as presented here for their fractionalized counterparts.

In conclusion, we have shown that Brillouin and Raman scattering resonant with the Mott gap are useful probes of spin-liquid physics, and in layered systems can potentially be used to detect chargeless topological surface states that cannot be seen with conventional surface probes such as STM and ARPES. Though we have focused calculations on a 3D model QSL phase, the qualitative lessons apply more broadly and may prove useful in studying protected surface states arising in other strongly-correlated systems.

\begin{acknowledgments}
	{\it Acknowledgements.}
	We acknowledge helpful discussions with K. O'Brien, D.L. Kovrizhin, R. Moessner, J. Rau, I. Rousochatzakis, A. Smith, and Y. Sizyuk. BP and FJB. acknowledge the hospitality of the Perimeter Institute. The work of BP was supported by the Torske Klubben Fellowship. JK is supported by a Fellowship within the Postdoc-Program of the German Academic Exchange Service (DAAD). NP acknowledges  the  support from NSF DMR-1511768. FJB is supported by NSF DMR-1352271 and Sloan FG-2015- 65927.
\end{acknowledgments}

\bibliographystyle{apsrev}
\bibliography{Refs_PRL}

\begin{thebibliography}{71}
\bibitem[{hfo()}]{hfoot}
\bibinfo{note}{The magnetic fields producing this value would be on the order
  of one Tesla in the iridates.}
\end{thebibliography}

\onecolumngrid
\appendix

\section{Supplementary Material} 

The isolation of spectra representing distinct representation of symmetry classes, known as symmetry channels, often involves taking linear combinations of multiple observable spectra. 
The particular linear combinations required for the symmetry group $C_{4v}$ of layered cuprate superconductors has been tabulated in Refs. \onlinecite{shastry91} and \onlinecite{sulewski91}. All of the channels, symmetric and anti-symmetric, can be found by taking scattering spectra at a few polarization combinations, often involving circularly polarized light. The low symmetry of the hyperhoneycomb model in a magnetic field makes this case more involved and we therefore present here an example set of observations that lead to the $[ac]$ channel discussed in the main text. The result applies for any antisymmetric channel $[\alpha \beta]$ with no symmetry assumptions except that $\alpha$ and $\beta$ be perpendicular, although the more general case follows from this one trivially.

Near resonance the light-scattering operator is a $3 \times 3$ matrix in cubic polarizations $\alpha,\beta = a,b,c$ \cite{perreault15}.
\begin{align}
R = \boldsymbol{\epsilon}_{\text{in},\alpha} R_{\alpha \beta}  \boldsymbol{\epsilon}_{\text{out},\beta}^*,
\end{align}
where the complex conjugation occurs only for photon-creation and is important only for polarizations with a circular component.
We allow ourselves to use labels for other vectors in place of $a,b,c$ to represent $R$ in polarization channels other than these. One example is the coordinates rotated by $45$ degrees: $\hat{a}' = ( \hat{a} + \hat{c} )/\sqrt{2}$ and $\hat{c}' = (-\hat{a} + \hat{c} )/\sqrt{2}$. These can be decomposed in terms of the cubic coordinates as
\begin{align}\label{R45}
2 R_{a'a'} &= R_{aa} + R_{cc} + 2 R_{(ac)} \\
2 R_{c'c'} &= R_{aa} + R_{cc} - 2 R_{(ac)},
\end{align}
for instance, where $2 R_{(ac)} = R_{ac} + R_{ca}$ and $2 R_{[ac]} = R_{ac} - R_{ca}$. Using Eq.~(\ref{I}) from the main text we can use Eq.~(\ref{R45}) to decompose the spectra in the rotated coordinates terms of ones in the cubic coordinates leading to
\begin{align}\label{IR}
4I_{a'a'} = I_{aa} + I_{cc} + 2I_{aa,cc} + 4I_{aa,(ac)} + 4I_{cc,(ac)} + 4I_{(ac)},
\end{align}
for example, where the mixed polarizations $I_{A,B} = 2\pi \int d\omega e^{i\omega t} \braket{[ R_A^\dagger(t) R_B(0) + R_B^\dagger(t) R_A(0) ]/2}$ are not directly observable and must be inferred from other measurements. 
We additionally define the left and right polarization vectors $\sqrt{2} \hat{r} = \hat{a} + i\hat{c}$ and $\sqrt{2} \hat{l} = \hat{a} - i\hat{c}$. Note that this assumes that both incoming and outgoing light travels in the direction normal to the $a-c$ plane. One can infer the desired spectra using only linearly polarized light by making measurements along additional non-orthogonal directions, but we do not report on those results here since back-scattering is the typical experimental setup and allows for cleaner results.

Using decompositions such as Eq.~(\ref{IR}), as well as the more trivial relation $I_{ac} + I_{ca} = 2(I_{(ac)} + I_{[ac]})$, the desired polarization combination can be related to the observable ones by
\begin{align}
4 I_{[ac]} = (I_{ac}+I_{ca}) - (I_{a'a'}+I_{c'c'}) + (I_{rr} + I_{ll}).
\end{align}
If we have a symmetry of the Hamiltonian taking $c\to -c$, as in the low-energy theory considered in the main text, or $a \to -a$, then we have that $I_{a'a'} = I_{c'c'}$ and $I_{rr} = I_{ll}$ so that one need only measure four independent spectra.

Finally, we note that in practice one may already be able to see the low-energy behavior described in the main text in directly-observable spectra for which the symmetric bulk spectrum is suppressed at low energies. Such is the case for the spectrum $I_{ab}$, for which the symmetric part $I_{(ab)}$ was shown to have a vanishing bulk response below $4J$ in the absence of a magnetic field \cite{perreault15}.

\end{document}